\documentclass[12pt]{article}

\usepackage[dvipsnames]{xcolor}%this package strictly necessary and before any other package
\usepackage{amsmath}
\usepackage{amssymb}
\usepackage{dsfont}
\usepackage[english]{babel}
\usepackage[utf8]{inputenc}
\usepackage{times}
\usepackage{graphics}
\usepackage{graphicx}
\usepackage[T1]{fontenc}
\usepackage[official,right]{eurosym}
\usepackage{url}
\usepackage{eso-pic}

\title{From decision aiding to the massive use of algorithms: where does the responsibility stand?}
\date{}
\author{O. Bellenguez$^+$\footnote{Corresponding author, odile.bellenguez@imt-atlantique.fr}, N. Brauner$^\dag$, A. Tsouki\`{a}s$^\ddag$ \\ $^+$IMT Atlantique, LS2N, Nantes, France \\ $^\dag$Univ. Grenoble Alpes, CNRS, Grenoble INP, G-SCOP, Grenoble, France \\ $^\ddag$CNRS-LAMSADE, PSL, Universit\'{e} Paris Dauphine, France}
\begin{document}

\thispagestyle{empty}

\enlargethispage*{8cm}
 \vspace*{-38mm}

\AddToShipoutPictureBG*{\includegraphics[width=\paperwidth,height=\paperheight]{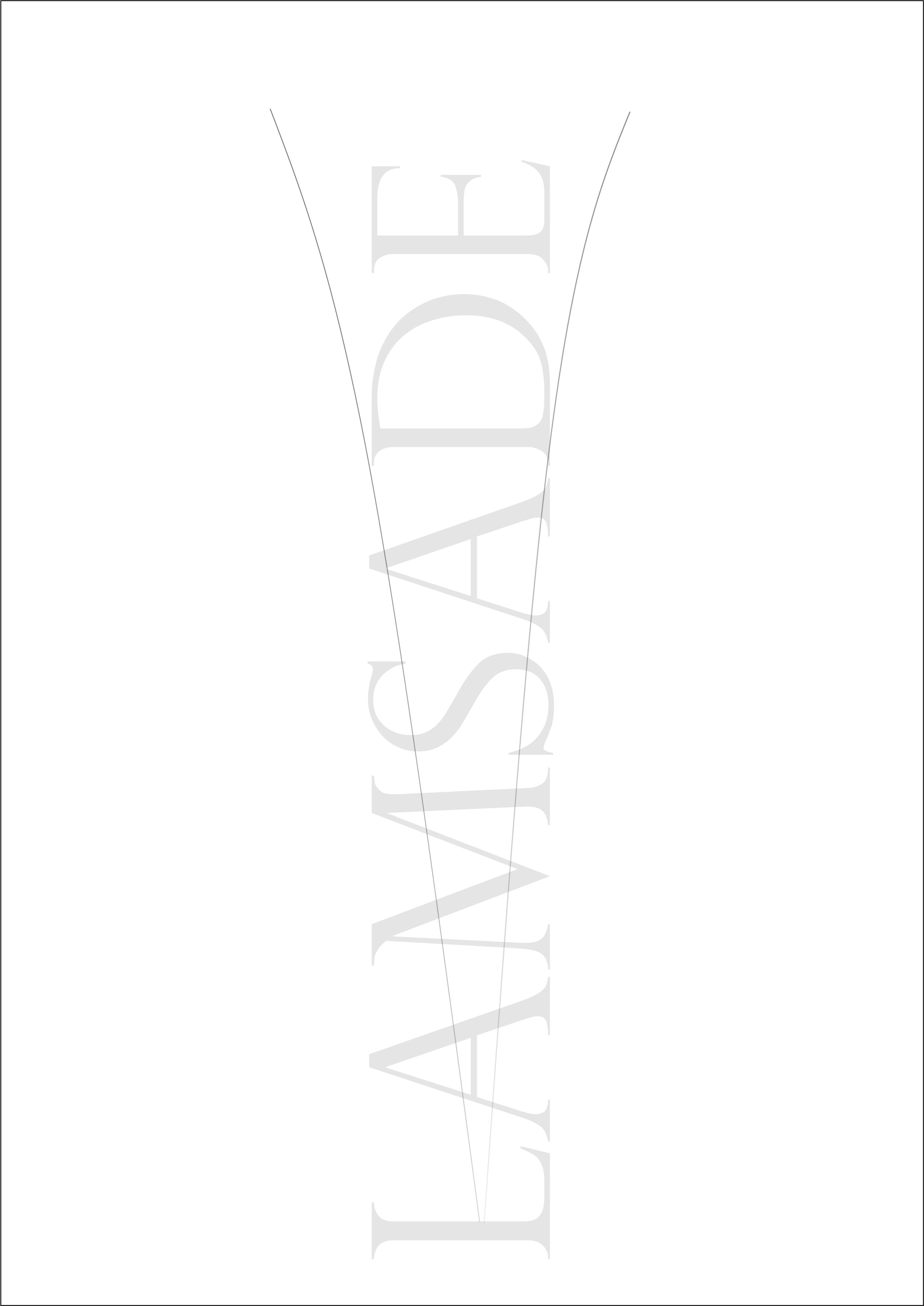}}

\begin{minipage}{24cm}
 \hspace*{-28mm}
\begin{picture}(500,700)\thicklines
 \put(60,670){\makebox(0,0){\scalebox{0.7}{\includegraphics{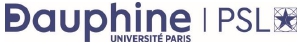}}}}
 \put(60,70){\makebox(0,0){\scalebox{0.3}{\includegraphics{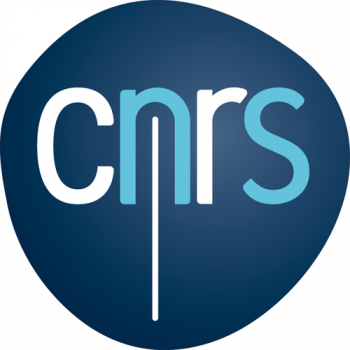}}}}
 \put(320,350){\makebox(0,0){\Huge{CAHIER DU \textcolor{BurntOrange}{LAMSADE}}}}
 \put(140,10){\textcolor{BurntOrange}{\line(0,1){680}}}
 \put(190,330){\line(1,0){263}}
 \put(320,310){\makebox(0,0){\Huge{\emph{408}}}}%XXX number of the Cahier: provided by Eleni
 \put(320,290){\makebox(0,0){June 2024}}
 \put(320,210){\makebox(0,0){\Large{From decision aiding to the massive use of algorithms: }}}
\put(320,190){\makebox(0,0){\Large{where does the responsibility stand?}}}
 \put(320,100){\makebox(0,0){\Large{Odile Bellenguez, Nadia Brauner, Alexis Tsouki\`as}}}
 \put(320,670){\makebox(0,0){\Large{\emph{Laboratoire d'Analyse et Mod\'elisation}}}}
 \put(320,650){\makebox(0,0){\Large{\emph{de Syst\`emes pour l'Aide \`a la D\'ecision}}}}
 \put(320,630){\makebox(0,0){\Large{\emph{UMR 7243}}}}
\end{picture}
\end{minipage}

\newpage

\addtocounter{page}{-1}

\maketitle

\textit{For the purpose of Open Access, a CC-BY public copyright licence
has been applied by the authors to the present document and will
be applied to all subsequent versions up to the Author Accepted
Manuscript arising from this submission.  https://creativecommons.org/licenses/by/4.0/} \\

\begin{abstract}
   In the very large debates on ethics of algorithms, this paper proposes an analysis on human responsibility. On one hand, algorithms are designed by some humans, who bear a part of responsibility in the results and unexpected impacts. Nevertheless, we show how the fact they cannot embrace the full situations of use and consequences lead to an unreachable limit. On the other hand, using technology is never free of responsibility, even if there also exist limits to characterise. Massive uses by unprofessional users introduce additional questions that modify the possibilities to be ethically responsible. The article is structured in such a way as to show how the limits have gradually evolved, leaving unthought of issues and a failure to share responsibility.
\end{abstract}

\section{Introduction}

The ethics of algorithms is increasingly the subject of much debates, without the questions being exhausted. First of all, it is essential to remember that this does not, or does not directly, involve technical aspects. In fact, all types of algorithms are likely to have unintended, undesirable or even unacceptable effects \cite{o2017weapons}. From decision-support approaches in Operational Research to Large Language Models, all advanced algorithms therefore deserve to have ethical issues considered as a necessary angle of reflection \cite{rapport2017comment}. This is why we will be talking about the ethics of algorithms in general.

But what is it really about? Ethics itself is concerned with what makes a life good and just, including the perpetual reflection on what makes a decision acceptable or even desirable in the light of higher dispositions, values or ends.
The concerned decisions  are made by subjects who are considered to be -at least partially- autonomous, in the philosophical sense of the term, voluntary and conscious.
In other words, subjects are responsible, i.e. in charge of answering for a decision, whether they have to justify it, amend it or bear the consequences. Responsibility is the other side of human agency. Agency that algorithms do not have.
Thus, responsibility cannot rely on an algorithm, in the sense that it is not philosophically autonomous, voluntary or conscious. The large literature on the need to keep a human in the loop clearly underlines the human responsibility that remains involved. The expression 'ethics of algorithms' is therefore a misuse of language, and encompasses questions of a different order that need to be distinguished \cite{tsoukias2021social}.

On the one hand, there are questions about the very technical means that will enable humans to retain their autonomy and exercise their full responsibility, since ultimately they will have to take responsibility for the decision, whatever help they receive from the algorithmic support in making it. Which is precisely why such tools are created: to support decision-making by gathering relevant information and exploring potential alternatives and consequences.

%revoir ordre
On the other hand, there is a central question that arises from the fact that these increasingly sophisticated tools are not neutral. The light algorithms provide on the decision is in fact guided by choices and performance measures constructed beforehand. The results or recommendations provided may also replace all or part of the analysis without being always explicit. However, these algorithms are produced by other humans, different from the users. These algorithms are designed after various choices have been made, choices which may be technical, but may also convey values or representations. Science is done trying to capture neutral knowledge. Nevertheless, technologies and scientific uses are never perfectly neutral. The lack of transparency around this non-neutrality is a major factor in the questioning in society as a whole, in the face of increasing consequences.
%{\color{blue}Today, this lack of neutrality of decision sciences and technologies is a major factor in the questioning in society as a whole, in the face of uses and their increasing consequences. }
With this in mind, one may wonder how responsibility is shared between the parties. For the purposes of this analysis, we will distinguish between two very large categories of actors: the decision-makers, who have the power to make the decision (with the help of an algorithm, and so are users) and carry in theory the responsibility; and the analysts, who receive the demand for aiding, so are expected to use methodological skills to design and develop the algorithm rather than competencies related to the decision process field, and are more and more questioned on their ethical responsibility. Obviously, all are in a given legal, social and economical context that has a great influence, this one coming in particular from the client, who wants the decision process to be assisted (and can or not be different from the decision-maker), the field experts, software editors and subjects impacted (users or customers).

The paper is organised as follows: in Section 2 we look back at the Operational Research literature, showing that questions relating to the ethical issues involved in decision support have been at the heart of debates for a long time and on several different issues. In Section 3 the emerging effects of the increasing automation of decision processes and decision aiding processes are discussed. In Section 4 we discuss the problems related to the massive use of decision support algorithms. In Section 5 we present an experiment conducted with students of a CNRS Summer School as far as the design of ethically responsible route planning services is concerned. We conclude summarising and identifying further research questions.

\section{Decision aiding}

Mankind has developed many methods of analyses or modelling of decisions, embedded in tools or devices, hoping to deal with the uncertainty that always surrounds decision-making. Operational Research (OR) in particular includes methods to tackle large classes of \textit{decision problems} (see \cite{colorni2024decision} for a formal definition) that are now used in lots of software tools. Such methods,  based on 'rational' tools \cite{meinard2019rationality,tsoukias2007concept}, are used since the first half of the twentieth century to assist professional decision-makers, especially for production, organisation, management or transportation.
This field has been considering the interactions with ethical issues for some time. %{\color{blue}
As \cite{wenstop2010operations} has summed up, since the 60's OR communities questioned several ethical aspects and limits of their practices, together with obligations that it reveals for OR experts as analysts.   Equivalent or complementary considerations also appear in the field of machine learning approaches \cite{wiener1960some, samuel1960some}. Nevertheless, current literature is far from dry and even if specialists admit that 'Although we may not be liable for what our clients decide using our advice and/or our tools, we are responsible for many (avoidable) consequences which can occur. We have a power and we need to use it with responsibility' \cite{bellenguez2023there}, there still is fuzziness around responsibility to explore.

\subsection{First steps}
The first point, widely identified by the pioneers of OR, relates to the fact that the algorithms initially designed to help decision-makers in a professional context obviously take into account the expressed needs of these decision-makers. In a decision support approach \cite{tsoukias2008decision}, the first stage is to understand the problem and build a representation of it. This stage directly involves the decision-makers to be assisted and will therefore integrate their requests first. But these requests are obviously not perfect. First, the way decision-makers express their needs could be partial, or could improperly paint the picture of their real expectations. %{\color{blue}
Second, as in the absence of a decision-support tool, the questions selected and the objectives to be pursued by the algorithm may therefore present blind spots, from an ethical point of view or not. Requests mainly target short-term productive goals. As \cite{churchman1970operations} stresses in particular, %,  this does not consider every stakeholders' will, but only a given utility, designed according to the decision-makers. Moreover, this author points that
representing a given decision for several players, to achieve a goal selected by the decision-makers, contradicts the Kantian obligation not to consider actors as means, but always as ends. It can also be seen as a problematic stance, lacking the necessary veil of ignorance \cite{rawls1971theory} that is a prerequisite for any form of Justice. More broadly, this orientation of expectations is nowadays increasingly criticised as not being socially acceptable. As a result, analysts are supposed to consider broader issues than just helping applicants.   %}

Furthermore, when a decision-support algorithm is put in place, its efficiency will have an amplification
effect in the particular sense of prior orientation of expectations.  However, the more effective it is on a given criterion, the greater is the risk of compacting the solution, leading to a reduction of (considered as) unproductive time, which in practice is used for recuperation, consultation or other purposes. As a running example, we illustrate the questions on routing problems (in italic) all along the paper: \textit{when a unique professional is looking to build manually a distribution route for her products, she tries to minimise the distance, but it will be difficult to explore all the solutions to find the one that suits best and reaches the optimum according to the considered criterion. Thus, there often still exists room at the end, in the sense that other criteria are not completely crushed. In that case, drivers can adapt their practice a little. On the opposite, if the demand is in this direction, an algorithm can very efficiently reduce the distance or the travelling time. But such orientation potentially creates work-related suffering for the driver}. This general intensification of working hours when focusing on immediate productivity is well documented \cite{gaudart2016activity}. \cite{bellenguez2020recherche} underlines how it could be linked with the use of optimisation approaches and not always detected quickly or not possible to correct. A crucial topic here is: the demand for an ``hyperproductive'' is an explicit demand from the client or has been introduced without being aware of it just because we adopted a certain objective function or a certain optimisation perspective? Ackoff \cite{ackoff1974social} cites a nice example where an optimised monetary reward function has been rejected by the workers concerned generating conflicts, while a free time reward function has been accepted with all parties being satisfied. Indeed the monetary function has been introduced only because the analysts were thinking only in terms of monetary compensations.

This alignment of algorithms with the stated objectives of the decision-maker, as they stem from particular interests and particular representations of the problem situation, partial and biased, also suffers directly from a lack of universalism as developed by \cite{churchman1970operations}, which cannot be seen as ethical, and which was identified during the initial questioning in this area. This has led to introduce the need to take into account the various stakeholders when designing algorithms (see \cite{ackoff1974systems, ackoff1987business}) that analysts have to be aware of, in order to lead properly the discussion. It shows just how important it is for the analyst to take into account different points of view and balance potential effects when formulating the problem to be addressed. Given the efficiency of tools based on scientific principles, part of the responsibility that used to rely on the decision-maker is shifting to the analyst, as a professional in the field of decision support.
% tu as ce que tu as demandé mais ce que tu voulais

\subsection{Analysts and normative ethics effort}
\label{axio}
Algorithms produce reproducible processes which, for the same input conditions, have to recommend the same result. Thus, they have to follow the rules that are prescribed and designed beforehand. This formal work is what makes the decision support system relevant for saving time and making practices more efficient in the general case. For this reason analysts are expected to put in place a process that respects justifiable and reproducible rules or principles, established long before real situations of use. If we focus on the ethical analysis in the strict sense of the term, the reflection is also carried out in advance of the usage phase, in order to consider the general case and set rules or principles to stick to. Thus, while the analyst takes on a part of responsibility and ethical vigilance,
this can only be done on a theoretical level, based on defined rules extracted from practice but not dedicated to each direct real problem situations. That is why it relates only to the level of normative ethics, which can be defined as a theoretical study of ethical behaviour to investigate questions regarding how to act, in a moral sense. This level of thinking has to be distinguished from applied ethics, that considers the practical aspect of ethical considerations and will be discussed in the following.

There are a number of points to bear in mind here. First of all, while the expressed request  may include a particular orientation coming from the decision-makers, the formulation of the problem may also be oriented by the representations of the analysts, who have their own visions of the context, and their own values, which are difficult to overcome. This obviously constitutes a limit to the supposed neutrality of the decision aiding process, and therefore of the algorithm that will be produced.

Secondly, this work of developing algorithmic treatments also has axiological limits, intrinsic to the scientific field: to be effective and valid, methods must be based on properties and hypotheses that must be guaranteed, which is not always possible in practice. Sometimes we may perfectly describe what we want to have as a procedure but do not know any technical way to achieve it.
%{\color{blue}
Considering Arrow's theorem \cite{arrow2012social}, we know for example that it is impossible to aggregate voters' preferences, respecting universality, unanimity, independence and absence of dictatorship. Thus, in practice we cannot for example build a voting system that will always produce a winner and that cannot be manipulated \cite{SATTERTHWAITE1975187}. Thus, we have to make trade-offs or choices: rely upon the Borda method, or on a two-round voting, but without guaranteeing the absence of manipulation, and this leads in raising ethical issues.

Sometimes, we have to face practical infeasibility: the complexity of the algorithm chosen could lead to prohibited computational time, or it requires data we do not have access to.  \textit{Consider the case of constructing a driver's tour. As the only exact approaches we know are exponential in time, it can take too long to find the best solution in some large real cases, so we focus on solutions close to the optimum, that have to be reached using relaxations or approximations that are based on techniques relying on the analyst's own expertise}. Those technical approaches and their limits can take a back seat to the ethical norms that we might have wanted to respect.

To these limits, we can also add the problem of the 'law of the instrument': When you have a hammer, it's tempting to think of everything as a nail \cite{kaplan2017conduct}. The analysts have acquired expertise in many classic decision problems, as well as in algorithmic approaches to solve them, so they will have a certain tendency to refer to this previous knowledge when answering a new problem. If analysts work on combinatorial optimisation, everything tends to be a combinatorial problem. \textit{When a decision-maker wants to build the 'best' vehicle routes, analysts often tackle the problem by looking for the routes with the shortest distance or minimum duration, which are classical objective functions to evaluate the 'utility' of a solution. It is much rarer to take into account the place where the driver wants to take a lunch break, despite the fact that this is an issue of interest and so could constitute an element of 'utility'}.

This point is in line with the expressed concern about taking stakeholders and consequences the analysts have to take into account during the analysis phase. Nevertheless, the tools are designed for generic cases and are based above all on the expected regularity. This directly prevents treatments to consider the uniqueness of each situation. This respect of arguable rules is the reason why we can speak of a reflection essentially at the normative - or theoretical - level, under the responsibility of the analysts, in order to frame as much as possible the results, but also leads to potential framing effects of what is called 'invisible technologies' \cite{berry1983technologie}. Thus, they have to question lots of dimensions we have just mentioned. But this cannot address every issues that may occur in practice, as we want to develop now.

\subsection{Decision-makers and ethics in practice}
Such generic treatment always becomes a source of potential undesirable effects, when circumstances change, without this being reflected in the input data, which transcribes the information that was thought necessary at the time of design. In real-life problem situations, there are always unforeseen events that occur and call for a rethink of the treatment, where theoretical ethics reaches its limits in the face of unbridgeable exceptions or contradictions. This is where practical ethics must take over.

However, in terms of practical ethics, something remains impossible to transfer to the analyst specially because it cannot be formalised with rules or capture with statistics. Indeed, the decision-maker (we previously define as the one who makes the final decision with the help of an algorithm), who is the problem expert %professional
in charge of the action to come, is by definition the person who has acquired something like \textit{practical wisdom}, in Aristotle's sense \cite{aristote1959ethique}, an on-the-ground knowledge of the problem. Unlike the analyst, these decision-makers have had recurrent experience of the problem situation for which a support tool is to implement. The decision-makers are therefore aware of a range of borderline cases, exceptions, difficulties and conflicts, that the analysts cannot all identify during discussions and model. The decision-makers, confronted with some of the affected stakeholders, such as the \textit{driver} in the given example, have already received feedback and noted possible difficulties. They have also already tried out different types of response and can assess many of the consequences, whether desirable or not. In a given situation, these decision-makers can rely on resources based on practical knowledge, not always theorised, but effective and adapted to the situation. And they mobilise this knowledge to arbitrate on the action to come in the decision moment: what might correspond to a form of practical ethics cannot therefore be fully transferred to the analyst upstream of use.

This leads here to highlight an initial line of responsibility sharing between the decision-maker, as an expert in the practical situation, and the analyst as an expert in decision problems. It can be argued that exchanges between these two parties enable mutual knowledge and information sharing. The fact remains, however, that dividing up responsibility necessarily complicates the analysis of the issues and the way they are taken into account, and even leaves certain cases or disputes outside the scope exchanges deal with, each one believing it to be within the supposed scope of the other, but in fact in an undefined area.

\subsection{Global issues}
%The analysis and consideration of stakeholders, which is discussed in the literature, also mentions wider issues, particularly at the social and ecological levels, for those directly affected but also for future sustainability (see for example \cite{brans2007ethics}). Evoking these issues echoes the ethical questions surrounding technoscience, amplifying the power of human action. Such an analysis obviously goes beyond the practical reflection that can be carried out by decision-makers on a daily basis, but also the conceptual ethics carried by analysts. In fact, as Hans Jonas\cite{Jonas79} shows, this level of questioning of what is at stake undermines all conventional ethical approaches.  What's more, in order to begin to sketch out answers, one would need to be able to rely on scientific knowledge dedicated to these identified fields of impact. At the scale of the creation of a single delivery tour for a producer, how is it possible to grasp in a relevant way the issues of the territories impacted by the traffic, or the ecological issues of carbon-based transport? It is reasonable to assume that this goes beyond the knowledge of the decision-makers and the analysts, leaving a vast territory that is difficult to address. This question, which is becoming more pressing with the spread of algorithms, will be taken up in part \ref{mass}.
The analysis and consideration of stakeholders, which is discussed in the literature, also mentions wider issues, particularly at the social and ecological levels, for those directly affected but also for future sustainability (see for example \cite{brans2007ethics}). Such an analysis obviously goes beyond the practical reflection that can be carried out by decision-makers on a daily basis. \textit{At the scale of the creation of a single delivery tour for a supplier, who is subject to an injunction of profitability,  how is it possible to grasp in a relevant way the issues of the territories impacted by the traffic, or the ecological issues of carbon-based transport? }It is reasonable to assume that this goes beyond the knowledge and power of the decision-makers. Some discussions \cite{reisach2016creation, ormerod2013operational} question it in the scope of what OR should address. Nevertheless, such issues are becoming more pressing with the spread of algorithms, without being solved. We argue that it is even becoming worst, because in fact analysts do not have real means of action. That high level issues will be discussed in part~\ref{mass}, as the effects of mass use are even more far-reaching, whereas decision support limited to a single decision-maker or single sector of activity, as mentioned in part \ref{auto}, primarily affects the players in the sector and their relatives (\textit{just as driver's relatives may be affected by the driver's working time or fatigue}).

\section{Automation}
\label{auto}
Despite any reservations, algorithms and decision-making supports may be a real time-saver in many context. In addition, the systematic treatments produced, based on scientific foundations, tend to suggest that they are easier to frame and justify in a rational way than human decision making. These are the reasons why they are more and more deployed, leading to the use of a unique algorithm for numerous decision-makers. On the one hand, more and more processes are being automated, and on the other hand, more apparently comparable decisions are being automated using the same algorithms. This has a number of effects on the responsibility of the actors involved, which are outlined in this section.

   \subsection{Decision process components}
Several taxonomies on automation have been proposed \cite{vagia_literature_2016}. It leads to distinguish three theoretical components of the decision-making systems: the first one is the \textit{moderator}, that supervises the decision-making process and makes it moving forward; the second one is the \textit{generator}, designed to analyse data and build feasible solutions; the last one is named \textit{decider} and is in charge of making the final decision, among feasible ones \cite{cummings_collaborative_2009}. When there is no automation, the \textit{moderator} corresponds to the moment a problem situation is identified (\textit{such as 'distribution tour is not satisfying'}) and turn into a clear question on a given perimeter (e.g. \textit{'How to build the driver's tour in order to serve every client in a day?'}). Concerning the \textit{generator}, when handcraft decision-making is done, it deals with the moment where different solutions are created and evaluated for comparison, which sometimes leads to reframe the first question, in case where no satisfying solution can be found. Finally, the \textit{decider} step corresponds to the election of a given solution, among discovered alternatives, for action.

Each one of these three components may completely rely on the human agent (without any automation) or on a full automation and so the human agent does not have the hand or any veto. Between these two extreme positions, the 'human in the loop' part could result in a system 'mixed but more human' or 'mixed but more automation', or it could be equally shared. As \cite{leitao2022human} discussed, the equilibrium  is far from being possible to reduce to the question of when to defer difficult decision steps to human sensitivity.

 \subsection{Normative difficulties}
These different components are more or less intertwined in a decision-making process that relies entirely on humans. To automate the process, each stage needs to be clarified, and many choices need to be made before use. The higher the degree of automation expected of the components, the more the analysts have to make prior trade-offs to automate processes, i.e. not to require any human intervention afterwards, and to depend solely on pre-established parameters. This parametrisation is often wider than optimisation methods' parameters, in a complete Automated Decision-Making System, that is a clearly channelling processes. Once done, the scope and format of the input data or the method of exploring solutions cannot be questioned again for each run, since this would mean repeating potentially the entire analysis and therefore completely disregarding the expected time savings. This automation and the resulting initial parameterisation may first be seen as a change of scale compared with the development of a targeted algorithm for a single decision, but there is in fact a change in the nature of the process which has many consequences.

Indeed, when the same algorithm has to be designed for many different uses - and therefore many different decision-makers, since keeping human in the loop is still in force - it would be necessary to integrate the expectations of all of them on all known cases. It soon becomes clear that this is impossible in practice, for a number of reasons. The first is linked to the inevitable disagreements between the decision-makers concerned on the way to model the situations addressed and on the principles to be followed in their practical implementation. No matter how much they know about each other or how much time they devote to building a compromise, convergence cannot be perfect, nor can it be perfectly transcribed by the analysts.

In addition, thinking through the compromises to be done in each case and for each problem situation would require time that is difficult to limit, and which increases drastically with the number of participants \cite{richard2020does}. Finally, it is impossible in practice to bring together all the users of a decision support system, who will be the decision-makers. When the system is created, there may already be too many of them, or they may be too dispersed, so they are only represented by pilot groups or \textit{persona}. Subsequently, the tool can also be made available to decision-makers who were absent, because they have not yet been recruited for the functions concerned, or because the scope of deployment is increasing.

Regarding the analysis, some of the subtleties specific to each decision are thus aggregated. Different ways of arbitrating cases can thus become undifferentiated, and analysed at a more general level. The approach adopted to analyse a very large scope is necessarily done at a higher level, not even close to the finesse required to integrate ethical issues perceived by the various players (such as \textit{drivers}). %The algorithmic rules are thus crushed into a unified approach that can be far from the particularities of each practice.
It is in fact often difficult for analysts to capture or bring out rarely formalised processes from the practice of a decision-maker, but this becomes worse when there are many of them, and when they have distinct habits, contexts or actors to consider.  The gap between prescribed work (or the description that can be given) and real work, which is never entirely reducible \cite{dejours2016evaluation} for anyone including the decision-maker, also carries an ethical dimension, which seems impossible to grasp, especially when the analysis remains too far from the ground. %multiple fields.

  \subsection{Practical experience}
These difficulties, which are specific to the analysis stage, lead to the production of highly standardised treatments, whose variations are often limited in the face of exceptions and special cases. The effects of this standardisation on work have been widely discussed since the work of \cite{berry1983technologie}. %More recently, the deskilling engendered by the automation of many professional decisions has also been highlighted in a wide variety of fields.
We can thus see various elements showing that the use of digital technology is changing the relationship to work and so directly influencing decision-makers on practice. As a result, practical experience of ethical issues is also changing. Despite the regular confrontation with each unique case, the space for reflection remains restricted by the previously framed process, which modifies the means of analysing the issues in stake, i.e. on the level of practical ethics that the analysts could not address.

Decision-makers, who are often kept in the decision-making process to act as ethical guarantors, are both guided and constrained by pre-customised parameters. On the one hand, the recommendations that have been constructed to save time and guarantee the stability of the decision encourage them to make a particular type of decision without mobilising a great deal of questioning. On the other hand, the formatted interface and pre-established rules can inhibit the creation of exceptions and differentiated treatments even when decision-makers might wish to do so. As a result, there is little room for in-depth reflection on the issues at stake, and little opportunity to acquire direct ethical experience. When moving from an untutored problem situation to the use of a decision support, we can even postulate that there is a \textit{loss of ethical sensitivity}. In the same way that we speak today of professional deskilling in the face of tools, we can underline a \textit{loss of ethical skill}, which tends to make the decision less sharp. From this point of view, the use of decision aids seems to be in direct contradiction with the practical experience highlighted by Aristotle in the ethical reflexion \cite{aristote1959ethique}. Experienced decision-makers gradually lose the ability to carry out a detailed analysis of the decision on their own, and more inexperienced decision-makers are unable to do so for the same reasons. In this way, we can speak of a global ethical deskilling, the risk of which is obvious. \textit{If all the transport departments in a given group are faced with the same route planning software, with functionalities based on a more or less aggregated representation, it will be difficult for each decision-maker to take account of the different social and cultural contexts in order to protect each driver from various prejudices, sometimes not perceptible via the tool}.

Over and above the individual limitations of direct confrontation with situations that a digital tool can only provide an incomplete representation of, there is also a collective issue in terms of ethical skill. Indeed, the decision-makers as a whole, aided by the support, apprehend the decisions to be taken only or mainly through the tool. They perceive the issues and potential solutions not through direct experience, but through the representation used during the design phase, which has partially fixed the dimensions and parameters.  However, when it comes to ethical issues, as elsewhere, many difficulties arise from primitive unthinking and emerging situations, as we said in the previous part. Because they are guided in their practices by the tool, decision-makers are unlikely to detect necessary changes in the model, whether these are linked to a gradual evolution in the decision-making environment or to the appearance of black swans \cite{taleb2008black}. Even when changes are detected by one or more decision-makers, they are not, or little, discussed, and are therefore too rarely the subject of a new formalisation or a revision of the model in place within the support. It leads to a great difficulty for new structured knowledge to emerge and be taken into account. On the example \textit{dealing with drivers' tour, we could say that despite the many testimonies, and the formal work now accumulated on the social isolation of lorry drivers, which has increased with the introduction of digital management systems, this is not the subject of direct adjustment levers in most of the software made available to decision-makers}.

\subsection{Supporting practical wisdom? }
Of course, this type of apparent lack in the tools is primarily due to their prior orientation, as we emphasised in the first part. Decision support tools are built around dominant representations and expectations, particularly in terms of optimising lead times and costs, which leave little room for other aspects of organisation. Faced with these incentives, which often relay productivist expectations, it is difficult for a decision-maker to maintain or develop an autonomous stance in integrating the different levels of challenge. Despite the desire to keep a human element in the loop, the tool leads to shape the way they exercise their responsibility. Here again, the decision-maker's responsibility appears to be subject to that of the analysts and the choices they make. However, as we said, the analysts are not directly involved in the decisions to be made \textit{in situ} and do not have the practical experience of the professionals mentioned above. %The resulting design cannot therefore have the finesse of the decision-makers' representations.

The grey area between the responsibility of some and that of others seems to have increased. Besides, it is not easy to remedy the situation. In order to prevent a form of deskilling, elements could be incorporated into the design to encourage the decision-maker to carry out a more in-depth analysis of the recommendation produced on different dimensions. However, to do this we would need to rely on data that is often qualitative in nature (\textit{isolation or fatigue, for example, are difficult to quantify}), or even difficult to qualify and transcribe because it is directly linked to human perceptions (\textit{unfairness between drivers}, for example).  In addition, decision-makers should undoubtedly be allowed to explore solutions outside the framework, in order to feed their experience as well as their capacity to create alternatives. From this point of view, the underlying paradigms, specific to time saving and the definition of expected efficiency, or the priority given to ethical issues, need to be reconsidered. A great deal of work is being done on flexibility and ways of giving the decision-maker back control at key stages in the process or in response to alerts that need to be defined. For more details on such interactive methods, we refer to the analysis proposed by \cite{meignan2015review}. Unsurprisingly, however, these aspects are often in conflict with automation, which is supposed to guarantee the stability and transparency of the process, as well as its speed. Besides, the construction of compromises of this nature is not obviously aligned with widespread dissemination, and should remain focus on a well-studied and limited perimeter, such as proposed by \cite{bebien2024ethical}.

Finally, it should be noted that any mechanism, such as interactive optimisation methods, designed to place decision-makers more firmly in their position of responsibility when using automated support systems (or just trying to avoid ethical deskilling), increases the occasional risk that these decision-makers' agency will not be in line with the ethical issues initially identified. The reason is because they might take a different approach to their responsibility than the one imagined by the analysts when they designed the proposed framework. The existence of such different perspectives cannot be handled quantitatively as we do in multi-criteria decision analysis. Any MCDA model should be considered on its turn under the real ethical dimension of the decision. \textit{If design is done to integrate different indicators on fatigue, or drivers' preferences while computing tours, a final user, who is the considered decision-maker, may prefer  to pay attention to other criteria according to a real situation.} Indeed, it is impossible to guarantee at the design stage an ethical framework for all the effects of using a decision support system in any cases. Moreover, to rely exclusively on the individual and professional responsibility of the decision-maker is also to run the risk of generating unforeseen or undesirable effects, despite an ethical vigilance, because of the delicate dimensions specific to decision aids, raised in the section \ref{axio}.

\section{Massification}
\label{mass}
Beyond automation and tool replication within a given structure, we see more and more digital tools that are being massively deployed. Many decision support tools now exist and are available to the general public. The aim is to give as many people as possible direct access to services. The platforms set up to make it easier to consult the population, book medical appointments, pay tax or apply to university, for example, are designed to democratise services, helping people to have new choices, participate to more debates. This is not just a change of scale. Such platforms are designed for a very large number of users, who will have a variety of uses, in very different contexts. What's more, when they are widely distributed, the cultural and social contexts to be considered can be completely different.
It should be added that the expectations to be taken into account cannot, of course, be discussed with all the potential stakeholders. There are too many of them, and many cannot be identified at the design stage. This ties in with the profound difficulties in establishing the standards to be included in the tool, as mentioned above.
However, this difficulty is not enough to explain the issues that are now widely denounced.

\subsection{Decision-maker at the helm?}
The algorithms proposed for recommending products or itineraries, for example, are based on decision support approaches. %, derived from explicit models or based on data.
Many people use them to save time on issues that do not necessarily fall within their professional perimeter. As in previous cases, they cannot be assumed to have sufficient knowledge on the axiomatic bases and performative effects of algorithms to grasp the issues at stake, but in addition here, they can no longer be assumed to have the practical experience to measure the ethical stakes of the decision to be taken. With little - or no - dedicated experience or knowledge of the potential impacts, users, who do not always see themselves as decision-makers whose responsibility is at stake, do not necessarily have the reference points to question their own use. Unlike professionals, they have \textit{a priori} no practical wisdom to draw on, and obviously no professional code of ethics to draw on when making their decisions.

In addition, professional decisions generally directly involve several players (service users, professionals involved, etc.), identified by the decision-makers. In the case of personal use, this is not systematic. More generally, there is not necessarily a circle in which the first direct issues are discussed, as was the case with the \textit{drivers} in the given example. What's more, the nature of the issues regularly raised is not the same. It is no longer a question of a professional decision for others that is potentially unsuitable and reviewable by the decision-maker who is trained in this issue (as a judge is capable of looking critically at an automatically suggested sentence, or just as \textit{a transport manager can review the planned distribution}).
Conversely, in the case of use by the general public, decision-makers generally do not have sufficient means of analysing the complex issues involved in their decision. Many of the issues involved are not directly visible or analysable. For example, until it had been in use for several years and various studies had been carried out, it was impossible for any individual to measure what is the result of the stacking effect of decision supports, but they are in fact numerous. Just as few examples, we may consider all the regional impacts on roads, cities or rentals, because the way we travel or do tourism drastically evolves \cite{cardon2019algorithmes}. It was only after a significant amount of news stories and formal studies that usage trends showed their impacts over a very wide circle. \textit{Only a huge number of people who consider relief routes proposed by navigation tools generates a stacking effect and so troubles for local residents, which eventually require costly improvements}\footnote{Navigation Apps Are Turning Quiet Neighborhoods Into Traffic, Nightmares, New York Times, 2017, \url{https://www.nytimes.com/2017/12/24/nyregion/traffic-apps-gps-neighborhoods.html}}.

On the other hand, there are effects which are detrimental to the decision-maker himself, threatening his autonomy in particular: excessive data collection, nudge, alternatives not presented, and so on. There may be various reasons for this, which we will mention below, but these problems cannot be directly questioned by the decision-makers, since it is precisely their own agency and capacity for analysis that are reduced. Not only can we therefore establish that decision-makers' responsibility is \textit{de facto} limited in the context of use, but in fact we could even go so far as to question the term 'decision-maker' itself for these users.

\subsection{Specific Design issues}
During the design phase, several points should be noted. As mentioned, not all 'decision-makers' can be heard, and not all their potential uses have been identified. As a result, analysts cannot rely on a clear, stable and unique definition of needs as in a professional context, or on the formalised expertise of a decision-maker with whom they can interact. The analysts must therefore understand the need and refine the solutions proposed by other means. However, a specific advantage of mass distribution lies in the number of uses for which data can be collected, enabling both an analysis of the behaviours and a more detailed description of the decision-making environment. \textit{In the case of route planning, for example, it is possible to understand habitual journeys, as well as establishing the traffic conditions that need to be taken into account when calculating journey times in real time}. There is a key issue here in terms of the effectiveness of the service offered, which in a sense justifies this functionality in the name of the decision-maker's expectations. However, collecting data on the decision-makers themselves, and potentially influencing -or nudging- them, can be seen as detrimental to them at the same time.

More generally, analysts are faced with a poorly defined problem and an infinite number of potential uses. They are therefore faced with the challenge of finding ways of setting a framework within which to develop the search for alternatives, and at the same time curbing any excesses that might arise. Use is therefore more and more framed by limited parameters, by possible prohibitions, or by suggestions to guide thinking (\textit{the driver cannot choose the planned speed of his vehicle, for example}). This serves both to simplify day-to-day use and to limit the risks of hijacking or misuse.
The algorithms are also designed for standard everyday use, allowing users to search standard solutions. \textit{Decision-makers may search for a shorter route, for example, but do not take into account additional side dimensions (driver fatigue, scenic beauty, quality of stops, local residents affected by traffic, etc.)}.

To make the algorithms effective on such classical criteria (distance and time usually), \textit{secondary objectives} are used. We call secondary objectives any measure that is used to guide the algorithm or make intermediary steps run in a given direction, \textit{such as gathering as much data as possible or clustering paths into limited sets to find the shortest path more quickly}. These secondary objectives are not necessarily aligned with the original need of the user \cite{wiener1960some}. In extreme cases, this can even be prejudicial, for example if the use of data to infer behaviour reveals elements of private nature.
Analysts are therefore in direct contact with palpable ethical issues here, but reducing or redefining secondary objectives often directly threatens the response to the primary objective, or the safety of use itself, and it is therefore generally a utilitarian trade-off that is made, within the framework defined by the laws. This arbitration is based on the use at the centre and the assumption of these supposed superior positive effects, which should in fact also be questioned for a real ethical analysis.

What's more, this ethical questioning, which relies on the analysts, is not at the same level as the question of use itself. The aim here is not only to measure the direct impact of the decision and the action to come (\textit{does the proposed road meet the expected safety and relevance criteria?}), but to assess the impact on the knowledge and agency of the decision-makers in their decision-making, or to assess the overall effects (\textit{generating unplanned traffic on unsuitable roads, for example, leading to regional development needs}). Whether we are talking about the psychological effects on the decision-maker or on other impacted users, or whether we are talking about social effects (\textit{occupation and use of land}) and environmental effects (\textit{pollution, soil artificialisation, etc.}), it appears that analysts are also faced with issues beyond their technical and theoretical skills, for which they have little or no training. Like decision-makers, they therefore often have a construction based on dominant representations. Such general representations only change with a long delay, as when dedicated research establishes the extent of the impacts and a new knowledge is disseminated.

In theory, we would like all these tools to focus on the common good, going beyond the immediate interests of the decision-maker or any provider, but this is a political issue. Analysts are neither trained or skilled nor mandated to deal with political issues. The fact that we are collectively aware of a large number of shortcomings or inconsistencies in our expectations does not allow us to establish a clear public response. Thus, the practical issues are delegated to the level of analysis, in the hope that technical skills alone can produce the fair society we expect. However, this does not take into consideration the fact the analysts on their turn are subject to the same questions and controversies, do not necessarily hold a common vision about the society and what a fair society should be. %that overlooks the fact that the same questions and fault lines run through the circles of technical professionals as they do through society as a whole.
Expecting that a technical solution will settle a political controversy means concealing the social controversies behind it and the political responsibilities of who is supposed deciding. At the same time any results produced using such technical solution will have political consequences, for which the analysts become \textit{de facto} co-responsible, as if they had to assume a political mandate that once again they do not have.

\subsection{Principles and regulation}
Work on the regulation of massive algorithms has therefore emerged, with several objectives. First of all, there is an important need to regulate their use in order to avoid piracy and misappropriation with serious consequences. This expectation is generally considered to exceed the loss of autonomy for the decision-maker, and is therefore imposed on analysts. In addition, there is a certain consensus on restricting the means of action of the analysts themselves, prohibiting the harvesting of certain types of sensitive information or other problematic practices. Some of these elements then fall within the scope of detailed legal regulation, dedicated to the protection of players from individual, commercial or other harm. The AI Act \cite{EU_AI_Act}
is the most recent example trying to deal with that.

However, there are also a number of principles, which can be adapted by field of application, by territory, or at the discretion of the analysts. Here we could meet what early OR questioning reveals. This creates a level of ethical reflection that is placed above the analysis in order to represent the political issues and global expectations that frame the technical orientations analysts may consider. Since the last years, we can find numerous works on AI ethics principles. In \cite{floridi2022unified}, authors highlight five general principles commonly used: beneficence, non-maleficence, autonomy and justice - that are already introduced in bioethics - and explicability. As \cite{munn2023uselessness} has established, there is an obvious risk of ethical washing, as these principles are often generic and difficult to define without conflict with other values. There is therefore a risk that everyone will adopt the definition that is easiest for them, without really addressing the underlying issues. One might thus think that this Principalism, which is being put in place with different variants, is of very limited use in practice. However, this global approach, potentially involving society as a whole, offers at least two different avenues: As it has been done for proven and unacceptable dangers, it is entirely possible to feed into legislation formally prohibiting certain practices, and thereby guiding or framing the work of analysts.

In addition, it is not uncommon for analysts to find themselves faced with ethical issues for which they are in the front line in the construction of models, without having the relevant resources or knowledge to mobilise. In the same way as professional decision-makers require a level of reflection of the order of practical ethics, it is conceivable that analysts will gradually develop professional expertise, enabling practical ethical reflection, dedicated to massive algorithm design. To this end, the principles argued in the literature are an essential knowledge base to be mobilised and confronted with 'on-the-ground' reality during modelling. In the same way that the medical profession trains its staff and promotes its own ethical reflection, albeit in relation to the conceptual knowledge on which it is based, the training of analysts cannot do without this first level of general principles, to be constantly developed and questioned in practice. Just as an example, keeping non-maleficence as a core goal while developing any applications may prevent analysts from sludging users.

This is a necessary first step: for analysts to understand and appropriate the principles under discussion, and the reasons why this vigilance is necessary in the face of risk. This brings us back to the establishment of ethical principles, and the training of analysts, which were already being discussed by the pioneers of OR \cite{ackoff1974social}. Nevertheless, massification is raising far greater issues, and at a speed never imagined before. The use of algorithms, like all technoscience, has deeply altered human power to act, amplifying it drastically. The stacking effects of individual actions we already mentioned, and the unpredictable long-term consequences make the individual subject who is supposed to be 'responsible for his actions' an inoperative framework for analysing responsibility. Months after design and analysts' work, it is now possible for anyone to use their phone to plan a journey thousands of kilometres away, then to rent a place, adding to all the behaviours that are changing cities and drastically increasing property prices, but also pollution, creation of roads or long-haul routes, and ultimately changing the behaviour of everyone, even if they are not direct users \cite{cardon2019algorithmes}.

This is why the reflection and structuring of expectations at institutional level, the numerous academic studies and the potential increase in analysts' skills are not enough. The ability to contain the effects is out of all proportion to the scale of those effects. As \cite{Jonas79} has widely argued, in the face of this unprecedented power to act, the foundations of classical ethics are being called into question. \textit{The Imperative of Responsibility} refers in particular to the skills and knowledge to be mobilised, in addition to ethical reflection itself. Here, we have mentioned individual psychological issues, societal issues and environmental issues. It is therefore in each of these spheres that knowledge is needed to approach the possible consequences and thus channel human power to act. And while analysts are in the front line with their own knowledge of models and algorithms, they cannot be the only ones to take the lead and it is certainly far from providing a complete respond to give them the sole mandate to arbitrate on so many choices, with such far-reaching impacts.

Analysts are directly involved when designing methods, because they choose, willingly or not, the non-neutral way they encapsulate principles or standards in algorithms. The fact that they have to answer for their decision on the way to make it enforces to think about how to justify and amend procedures if needed. Nevertheless, uses to come do not involves them directly and cannot be fully imagined. That is why, users - decision-makers - also always bear a part of responsibility. But it can't be the only responsibility of each decision-maker to carry out on his own the full analysis every time an algorithm is used, since their practices are framed by previous choices and norms.

Even if it is difficult to define the perimeter of each actor's direct responsibility, as we have tried to do here, the uses and their consequences are in fact the responsibility of all those involved, in an attempt to grasp the whole picture and the issues at stake. So, it is essential for everyone to capture the stakes and try to address these issues on each given perimeter. It means for each potential use trying to explore, to imagine the worst, to call on knowledge that is useful for understanding and anticipating, trying to fully exercise one's responsibility for future actions, since the moment we are using such powerful tools.
But any design, any use are also coloured by habits, cultures and the whole environment, and we know that consequences also come with interactions and stacking effects to grasp in the whole picture. So, at the end, it relies on the global society. As the consequences come with the collective, it is first of all at this level that limits need to be found and where individual responsibility is diluted.

\section{Case study}
Trying to increase awareness on issues and how it is linked with scientific and technical considerations, a case study, built together with Christine Solnon, has been proposed to all the participants of the Summer School dedicated to responsibility of algorithms, that held in 2023 in Aussois (FR)\footnote{École th\'ematique du CNRS, ``Responsabilité des Algorithmes'', Aussois, 2023, \url{http://gdrro.lip6.fr/?q=node/297}}. Based on their personal knowledge on navigation tools (as Waze or Google maps), participants were invited to think about what could be a new tool with deeper ethical considerations. Another way to say it: The idea was to work on possible ethical questions to tackle in the analysis and design phase of a massive tool. This study has begun by discussing the article written by Cardon and Crépel \cite{cardon2019algorithmes} who mention many dangers and impacts of navigation tools on territories, behaviours and stacking effects. %Four different groups spent several hours working on the project before sharing their reflection.

Navigation tools are based on well-know approaches aimed at solving the shortest path problem under different settings of available information and type of demand. From a technical point of view the problem is extensively studied \cite{braekers2016vehicle}. Results around temporised-graph are to be considered for a real-time computation of travelling time \cite{fontaine2023exact}. The case study focused on additional questions to consider around impacts, and  the reasons for which responsibility could be hard to carry during the design of a tool. In the following we present different issues discussed during the exercise or after.

\begin{enumerate}

\item A first issue deeply discussed was dedicated to data collection, that could threaten privacy. Alternatives have been explored, using satellite images or phone signal aggregation to identify traffic jams. This raised a big question on the likely cost of balancing a lower level of individual data collection, if chosen.

\item Many exchanges have been around the stacking effects and impacts: in particular forced urban development to face traffic and environmental pollution coming from vehicles. Many solutions were explored to limit or nudge individual behaviours, by enforcing users to car sharing, co-modal travel, or shared-use vehicles. Channelling the possible routes to prohibit passing through high-density areas, or in front of schools and other sensitive locations, was also an option considered, which could be of interest to certain cities currently disrupted by traffic changes.

\item The wider issue of limiting travel as a whole in order to reduce pollution and energy consumption was also explored.
Most important principles or values to consider were obviously not the same for every participants, and so could change technical choices to make. Debates has made appear many social considerations behind travelling assistance, and have largely shown how algorithms design actually involves analysts as full-member citizens that they are.

\item A last point is related to the technical dimensions studied in computer science curricula. An important objective is to consolidate the understanding of shortest path algorithms (A*, Dijskstra) by modifying the problem and studying the properties that are or are not preserved. These modifications are the addition of the time parameter when crossing an arc, which does not allow to preserve the optimality property of the sub-problems. Then, the FIFO (First in, First out) property, which is natural in practice since it indicates that entering a route later means arriving at the exit later, restored this property but obliged the students to understand the algorithm and its proof and to adapt them.

\end{enumerate}

The discussion between these technical questions and the ethical issues led to very reach debates. It allowed to make appear how technical and ethical consideration could be intertwined and it increased awareness of participants on hidden political challenges. Discussions have also brought the deskilling question through difficulties Inuits now have with orientation because of such tools \cite{aporta2005satellite}. Finally, the discussion also gave very interesting avenues on how to extend such exercises to training courses for computer scientists and engineers, leading them to higher critical thinking. This study is now also offered to computer scientists as part of technical courses on shortest path algorithms (such as in Operations Research courses in Universit\'{e} Grenoble Alpes or in a course on Algorithms and Artificial Intelligence at INSA Lyon.)

%{\color{red}<NOTE> .}  %(see $https://moodle.caseine.org/mod/resource/view.php?id=61476$ for more details about possible scenarios, students ideas' on stakeholders, challenges, and questions on the way to balance positive and negative issues).
% in Operations Research courses in Université Grenoble Alpes or in a course on Algorithmes and Artificial Intelligence at INSA Lyon.
% Programme à Grenoble : https://formations.univ-grenoble-alpes.fr/fr/catalogue-2021/licence-XA/
%licence-informatique-IAI7UC15/parcours-informatique-generale-3e-annee-KI8WW05W.html
%(je le donne aussi en RO avancée à l'ENSIMAG mais c'est aussi l'UGA...)
% Il faudrait confirmer le nom du cours avec Christine
% Lien vers la page qui recence les ressources de cours sur éthique et algo. https://moodle.caseine.org/course/view.php?id=1007&section=2

\section{Conclusion}

Although algorithms and decision support systems are valuable allies for fast, reproducible and effective decisions, they may also have deep harmful effects, direct or not. For that reason, it is an important question to search for ways to ensure liability. If humans are the only beings capable of ethics \cite{Jonas79}, we discussed different limitations that decision-makers and analysts are facing to assume full responsibility, showing specifically how modelling is not just a technical question. In fact, we all have to assume responsibility as users and stakeholders, since massification results in cumulative effects that are far from being foreseeable. An important point concerns values or principles to apply in a collective manner. These depend on the whole society, at a political level, so debates and responsibility are to consider at this given level.
There may be issues and potential impacts on individual critical thinking and agency as well as on wider external perimeter, but the ability to be responsible does not increase accordingly to this impacted perimeter, for any single actor.

Global knowledge is currently increasing about potential harm, far beyond what first OR practitioners knew, and confirms that awareness of analysts is mandatory \cite{bellenguez2023there} but this awareness alone does not ensure a better result. Even the global awareness, more and more developed, that leads to elaborate principles, guidelines and thresholds, does not ensure harm or that any detrimental effects will disappear. We are facing creations with a level of impact bigger than recourse possibilities. In one word, we are deploying uncontrolled devices to unprepared uses.
In view of this, it is not only necessary to discuss the ethics of algorithms, or of the people who build and use them, but to accept the fact that we need to integrate this reflection, this perpetual questioning, at every stage, and on an ongoing basis. As we said, ethics includes the perpetual - individual and collective - reflection on what makes a given decision acceptable or even desirable in the light of higher dispositions, values or ends. So, the moment a process becomes automatic, massive and blindly used (\textit{i.e.} without any further reflection), there is no more ethics.

\section*{Acknowledgements}

Authors want to deeply thank every participants of the Summer School on Responsibility of algorithms in 2023, supported by the CNRS, GDR ROD and RADIA. The discussions that took place greatly enriched the reflections that preceded the preparation of this article. Authors also would like to address a particular thank to Christine Solnon for her great contribution to the case study and discussions, and Marc-Antoine Pencol\'{e} who has read an early version of this paper and provided precious advises. The last author acknowledges the support of two CNRS-MITI 80-PRIME grants supporting his research. Obviously, authors remain the only ones responsible for the contents of this essay.

\bibliographystyle{plain}
\bibliography{bibliomassiveethics}
%\bibliography{../../../complete-bibliography,../../../temp}

\end{document}